\documentclass{amsart}
\newcounter{mynr}
\newtheorem{thm}{Theorem}[section]
\newtheorem{cor}[thm]{Corollary}

\newtheorem{prop}[thm]{Proposition}
\newtheorem{defn}[thm]{Definition}

\numberwithin{equation}{section}

\newcommand{\abs}[1]{\left\vert#1\right\vert}

\newcommand{\Real}{\mathbb R}
\newcommand{\eps}{\varepsilon}

\newcommand{\E}{\mathcal{E}}
\newcommand{\Sc}{\mathcal{S}}

\newcommand{\ket}[1]{\left\vert #1\right\rangle}
\newcommand{\bra}[1]{\left\langle #1\right\vert}
\newcommand{\parket}[1]{\vert #1)}\newcommand{\parbra}[1]{( #1\vert}
\newcommand{\brkt}[2]{\left\langle #1 \vert #2\right\rangle}
\newcommand{\parbrkt}[2]{\left( #1 \vert #2\right)}
\newcommand{\braket}[3]{\left\langle #1 \right\vert #2\left\vert #3 
\right\rangle}
\newcommand{\schrod}{Schr\"odinger\ }
\begin{document}
\setcounter{mynr}{0}
\input epsf
\title{Time-Energy coherent states and adiabatic scattering}%
\author{J.E. Avron, A. Elgart, G.M. Graf, L. Sadun}
\address{J.E. Avron, Department of Physics, Technion, 32000 Haifa, 
Israel\newline\indent
A. Elgart, Jadwin Hall, Princeton University, Princeton, NJ 08544, 
USA\newline\indent
G.M. Graf, Theoretische Physik, ETH-H\"onggerberg, 8093 Z\"urich,
Switzerland\newline\indent
L. Sadun, Department of Mathematics,
University of Texas, Austin, TX 78712, USA}
\email{avron@physics.technion.ac.il, aelgart@princeton.edu,\newline 
$\left . \right .$ \qquad\qquad\qquad\qquad 
gmgraf@itp.phys.ethz.ch, sadun@math.utexas.edu}%

\thanks{This work is supported in part by the ISF; the Fund for 
promotion
of research
at the Technion, and the Texas
Advanced Research Program and NSF Grant PHY-9971149. The authors 
enjoyed the 
hospitality of
the ETH during various stages of this work.}%

\date{\today}%
\begin{abstract}
Coherent states in the time-energy plane provide a natural basis
to study  adiabatic scattering.  We relate the (diagonal) matrix 
elements of the scattering matrix in this basis with the frozen 
on-shell scattering data. We describe an exactly solvable model, and 
show that the error in the frozen data 
cannot be estimated by the Wigner time delay alone. We introduce the 
notion of energy shift, a conjugate of Wigner time delay, and show 
that for incoming state $\rho(H_0)$ the energy shift determines the 
outgoing state.  
\end{abstract} \maketitle
\section{Introduction}

Scattering from a slowly changing scatterer is described, to leading
order, by a {\em time independent} scatterer frozen at the scattering
time \cite{thirring}. Although this seems like stating the obvious, it
turns out that in trying to make precise how accurate this
approximation is, one encounters both conceptual and technical
difficulties. Our aim is to describe these difficulties and explain
how they are resolved.

One conceptual difficulty is to understand what the frozen S
matrix---a function of energy and scattering time---means.
Strictly speaking, a function of both time and energy is in
conflict with the uncertainty principle.   A wave that is sharp in
energy will have an ill-defined scattering time and conversely, a
wave with a well-defined scattering time is ill-defined in energy.
What, then, is the meaning of the frozen S matrix?

The resolution of this problem is related to the fact that the 
adiabatic limit  naturally leads to different parameterizations of 
time, and the right parameterization has small uncertainty.  
Specifically, the physical time $t$ will
parameterize  the intrinsic ``fast'' dynamics and has the usual 
time-energy uncertainty  $\hbar$. The slow variation in the external 
conditions will be parameterized by $s$. We refer to the latter as 
{\em epoch}. Since the epoch often plays a role of a parameter it is 
convenient to choose $s$ dimensionless. The two parameterizations are 
related by $s=\omega t$, with $\omega$ a slow frequency---the
adiabaticity parameter. The epoch-energy uncertainty then takes the 
form $\delta
s\,\delta e\sim \hbar \omega$ and so arbitrarily small in the  
adiabatic limit.

Coherent states provide a convenient basis to analyze the
semi-classical limit \cite{perelomov, hepp}. Semi-classical analysis 
is traditionally about the $\hbar \to 0$ limit, but is equally valid 
when $\hbar$ is fixed (and henceforth set equal to one) and $\omega 
\to 0$. Here we introduce
coherent states labelled by points in the time-energy plane, with
time being the scattering time. As we shall see, the frozen S
matrix approximates the diagonal matrix elements of the dynamical 
scattering
matrix in such coherent states. This reconciles the time-energy
uncertainty with the frozen scattering data. In a further step, matrix
elements of the frozen S matrix can be approximated by the on-shell 
data.

Another thorny issue that we address is when a description in
terms of frozen data is meaningful and how accurate it is.  The
question can also be rephrased as a question about the intrinsic
time scale relevant to scattering. If $\tau$ denotes this time
scale, then $\omega\tau$ is the error in the frozen data and 
$\omega\tau \ll 1$ characterizes the adiabatic regime. 

The Wigner time delay $\tau_w(E,s)$ conveys information about the 
time 
the particle spends near the scatterer. It is a function of the 
energy 
$E$ and scattering epoch $s$. It is tempting to 
hope that $\tau$ might be estimated by $\tau_w(E,s)$, but there is no 
compelling argument for doing so.
One cannot argue on the basis of dimensional analysis 
alone, since  $\dot \tau_w$ and $\sqrt{\tau^\prime_w}$, with dot a 
derivative with respect to the epoch and prime with respect to the 
energy, give additional and independent time scales. In fact, since 
Wigner time delay is a {\em comparison} of the arrival time at a 
faraway point, relative to the time of arrival in the free dynamics, 
it  is not even positive-definite. This  suggests that it cannot 
quite 
capture  $\tau$, which is more closely related to the ``dwell time'' 
near the scatterer. 

The way to determine $\tau$ is to consider the error in approximating 
the scattering data by the frozen data. The error is, to leading 
order, proportional to the adiabaticity parameter $\omega$. Since the 
error is, in general, complex, we identify $\tau$ with the absolute 
value of the error divided by $\omega$.
Calculating the error, to leading order in $\omega$, is 
no harder, and reminiscent of,
calculating the scattering in the lowest order of the Born
approximation. 

We shall see that, to leading order, the adiabatic time scale $\tau$ 
can be 
estimated from the scattering data and the derivative of the 
Hamiltonian $H$ 
with respect to the epoch, Eq.~(\ref{tau+e}) 
below, but not from the Wigner time delay alone. We show this by 
considering an exactly soluble model where the dynamical S matrix can 
be computed explicitly.

We introduce the {\em energy shift} operator $\E$. This is a measure 
of the energy change in time dependent scattering and is a natural 
dual of the Wigner time delay. As we shall see, in the case that the 
incoming state is $\rho(H_0)$,  the outgoing state is 
$\rho(H_0-\omega\E)$. In the adiabatic limit, the energy shift can be 
approximated by the frozen energy shift, which is related to the 
logarithmic derivative of the on-shell scattering matrix with respect 
to the epoch, Eq.~(\ref{on-shell-shift}). The energy shift then gives 
a handle on the exchange of energy \cite{asch,nakamura} and the 
pumping of charge in adiabatic scattering \cite{aegs}.

\section{Elements of scattering theory}
Scattering theory is a comparison of dynamics: One is the actual
dynamics generated by the time dependent $H(t)=H_s,\,(s=\omega t)$, 
the other 
is a fiducial dynamics generated by a {\em time independent}
Hamiltonian $H_0$. The Hamiltonian $H_0$ is the generator of dynamics 
for 
which there is trivial scattering and the S matrix is the identity.

The results of this section are true in general, without taking the
adiabatic limit $\omega \to 0$.  We shall assume that $H$ and $H_0$ 
admit good
scattering. Namely, we assume the existence of wave operators and
the unitarity of the S matrix. For explicit conditions on $H_0$ and 
$H(t)$ that
guarantee this see e.g. \cite{simon,yafaev}.

\subsection{The wave operator}

Let $U(t'',t')$ and $U_0(t'',t')=U_0(t''-t')$ denote the evolution 
from
time $t'$ to $t''$, generated by $H(t)$ and the time-independent
$H_0$ respectively. \begin{defn} The wave operators, based at epoch 
$s$, are
defined by the (strong) limit
\begin{equation}\label{wave-operator}
\Omega_\pm(s;H,H_0)=\lim_{t'\to\pm\infty} U(t,t')U_0(t'-t),\quad
(s=\omega t).
\end{equation}
\end{defn}
The existence of the limit, and the equation of motion imply
\begin{prop}
The dependence of the wave operator on the base point $s$ satisfy
the differential equation
\begin{equation}\label{omega-dot}
-i\omega\dot\Omega_\pm(s)=H_s\Omega_\pm(s, H,
H_0)-\Omega_\pm(s,H,H_0)H_0.
\end{equation}
\end{prop}

As we shall presently see, the notion of wave operator based at epoch 
$s$ is 
only interesting in the case of a time dependent $H(t)$.
\subsection{The frozen wave operators}
The frozen Hamiltonian
$H_s$ is \emph{time independent} so $U(t'',t')=e^{iH_s(t''-t')}$, in 
this case and
$\Omega_\pm(s_0,H_s,H_0)$ is independent of the base point 
$s_0=\omega t_0$. 
This follows from the existence of the limit in
Eq.~(\ref{wave-operator}) since $t'\to\pm\infty$ is the same as
$t'-t_0\to\pm\infty$.  To stress this we write $\Omega_\pm(H_s,H_0)$. 
 From Eq.~(\ref{omega-dot}) then follows the standard intertwining 
relation of time-independent scattering theory:

\begin{cor}
The wave operators $\Omega_\pm(H_s,H_0)$ relating the frozen
Hamiltonian at epoch $s$ and $H_0$ are independent of the base
point, and intertwine the two dynamics:
\begin{equation}\label{intertwine}
H_s\Omega_\pm(H_s,H_0)=\Omega_\pm(H_s,H_0)H_0.\end{equation}
\end{cor}


\subsection{The dynamical S matrix}

The (dynamical) scattering matrix based at epoch $s$ is defined by
\begin{equation}\label{S}
  \Sc_d(s;H,H_0)=\Omega^\dagger_+(s;H,H_0)\Omega_-(s;H,H_0).
\end{equation}
The S matrices based on different points in time are all related by
conjugation generated by the free evolution. Namely:
\begin{prop}
Suppose that the wave operators exist. Then
\begin{equation}\label{Sd0:Sds}
\Sc_d(s;H,H_0)=e^{-iH_0t}\Sc_d(0;H,H_0)e^{iH_0t},\quad
(s=\omega t).
\end{equation}
\end{prop}
This follows from
$U(s,t)\Omega_\pm(s;H,H_0)=\Omega_\pm(s;H,H_0)e^{-iH_0(s-t)}$.
Under a change of the reference Hamiltonian, say to the frozen
Hamiltonian $H_s$,
\begin{equation}\label{S-h:h1}
\Sc_d(s;H,H_0)=\Omega^\dagger_+(H_s,H_0)\Sc_d(s;H,H_s)\Omega_-(H_s,H_0).
\end{equation}

\subsection{The frozen S matrix} In the frozen S data the epoch is 
decoupled from 
time. As such it can also be studied using time independent methods,
which are normally quite powerful \cite{yafaev}. 
Its basic properties are in marked contrast with that of the
dynamical S matrix, namely:
\begin{cor} The frozen S matrix
\begin{equation}\label{frozen-s}\Sc_f(H_s,H_0)=\Omega^\dagger_+(H_s,H_0)\Omega_-(H_s,H_0)
\end{equation} is independent of the base point. It depends on the 
freezing
time parametrically through $H_s$.
\end{cor}

\subsection{The on-shell S matrix}

$H_0$ provides a basis that spans the Hilbert space of scattering
states. Let $\parket{E,j}$ denote the generalized eigenvectors of
$H_0$:
\begin{equation}\label{basis}
H_0\parket{E,j}= E\,\parket{E,j},\quad
\parbrkt{E,j}{E',j'}=\delta(E-E')\delta_{j,j'}.
\end{equation}
$E$ is the energy and $j$ labels the scattering channels.
 $\Sc_f$ commutes with $H_0$, by Eq.~(\ref{intertwine}), hence
\begin{equation}\label{on-shell}
\parbra{E,j}{\Sc_f\big(H_s,H_0\big)}\parket{E',j'}=\delta(E-E')\,
S_{jj'}(s,E).
\end{equation}
$S_{jj'}(s,E)$ is the {\em on-shell} scattering matrix. Note that
in the frozen Hamiltonian the physical time is decoupled from the
epoch, which now has been relegated to the role of a parameter. The 
on-shell scattering
matrix therefore is  not in conflict with the uncertainty principle.

\section{The energy shift} By taking the $s$-derivatives of
Eq.~(\ref{Sd0:Sds}) one gets
\begin{eqnarray}\label{dotS}
i\,\omega\dot\Sc_d(s)\Sc _d(s)^\dagger=
H_0-\Sc_d(s)\,H_0\,\Sc_d(s)^\dagger&=&[ H_0,\Sc_d(s)]\Sc _d^\dagger(s)
\nonumber \\ &=&
[ H_0,\Sc_d(s)-\Sc_f(H_s,H_0)]\Sc _d^\dagger(s).
\end{eqnarray}
This equation may interpreted as follows. If we think of $H_0$ as
the asymptotic observable associated with the outgoing energy,
then $H_{0,in}=\Sc_d(s)\,H_0\,\Sc _d(s)^\dagger$ represents the
asymptotic observable \cite{cycon} corresponding to the incoming
energy. This motivates calling
\begin{equation}\label{energy-shift}
\E_d(s)= i \dot \Sc_d(s) \Sc^\dagger_d(s).
\end{equation}
the operator of energy shift.

The energy shift vanishes for time independent scattering, as it
must. It gives a handle on changes in (certain) quantum states. By 
the functional
calculus applied to Eq.~(\ref{dotS}), for any function $\rho$:
\begin{equation}
\Sc_d(s)\,\rho(H_0)\,\Sc_d(s)^\dagger=\rho\big( 
H_0-\omega\E_s(s)\big).
\end{equation}
This is interpreted as follows: If $\rho(H_0)$ is the incoming
state, then the corresponding outgoing state is $\rho\big(
H_0-\omega \E_s(s)\big).$ 
The energy shift is a first order quantity in
the adiabaticity parameter and, as we shall see, it can be
approximated, to leading order by the frozen data. This then gives
a handle on the outgoing state $\rho$ to  first order in the
adiabaticity parameter.

\begin{prop}
The energy shift based on time $s$ is conjugate to the energy
shift based on  time zero
\begin{equation}\label{E-dot}
\E_d(s)= e^{iH_0 t}\E_d(0)\,e^{-iH_0t},\quad
(s=\omega t).
\end{equation}
\end{prop}
This follows directly from Eq.~(\ref{Sd0:Sds}) and
Eq.~(\ref{dotS}).
\section{The problem of adiabatic scattering}
The dynamical S matrix has qualitatively different properties from
the frozen S matrix: The dynamical S matrix has no freezing
time---It does not ``know'' when the incoming wave is going to hit
the scatterer. It does depend however, by conjugation, on a choice
of a base point. In contrast, the frozen S matrix is independent
of the choice of a base point and depends non-trivially on the 
freezing
time---The frozen scattering data for one epoch know nothing a-priori
about the corresponding data at any other epoch.

Matrix elements of the scattering matrix carry information about
the time that the wave is near the scatterer. For such matrix
elements, the adiabatic limit can be expressed in terms of the
corresponding frozen matrix elements. However, the introduction of
wave packets promotes the  epoch from  playing the role of a
parameter, to that of real, albeit slow, time. One then needs to
confront the uncertainty principle. We do that by considering
matrix elements between coherent states labelled by points in the
energy time plane.

\subsection{The Wigner time delay}  The Wigner time delay is defined 
in terms
of the on-shell scattering matrix. When this definition is
transcribed to the frozen, on-shell, S matrix it reads
\begin{equation}\label{wigner-delay}
\tau_w(s,E)=-i\, S'(s,E) S^\dagger (s,E).
\end{equation}
Prime denotes partial derivative with respect to the energy. With
this definition, the Wigner time-delay is a Hermitian matrix.

\subsection{The frozen energy shift}
For the frozen, on-shell, Hamiltonian one can associate a matrix
of energy shift  which is a natural conjugate of the Wigner time
delay:
\begin{equation}\label{on-shell-shift}
\E (s,E)=i\,\dot S(s,E) S^\dagger (s,E),
\end{equation}
where dot denotes derivative with respect to the epoch.

\subsection{Time scales}

The  frozen  on-shell S matrix defines several time scales. Among 
them:
$\tau_w$ and the (dimensionless) time scale $ \E^{-1}$. The
coherent states provide us with yet another time scale related to
the time-width of the coherent states.  One of the problems of
adiabatic scattering is to study the relation between these time 
scales and the  time scale $\tau$ such that
$\omega\tau\ll 1$ characterizes the adiabatic regime.

\section{Time-Energy Coherent states}
\subsection{The role of dispersion}
For a particle moving on the line, its energy and the time that it
crosses the origin are canonical coordinates.  One can therefore
construct energy-time coherent states in analogy with the usual
phase space coherent states. The explicit construction, however,
depends on the dispersion law. For linear dispersion the
construction is particularly simple.

Consider a classical particle with dispersion law $e(p)$ moving
freely on the line. The velocity of the particle is $e'(p)$ so the
time of passage through the origin is $t=\frac {-q}{e'(p)}$.
Time-energy  are (local) canonical coordinates since
\begin{equation}\label{canonical}
de\wedge dt=dq\wedge dp.
\end{equation}
The global aspects of the energy-time phase space can be
complicated. For example, for a free (massive) particle, with
quadratic dispersion $e(p)={p^2} $ the energy-time phase space is
made of two copies of the half plane $e\ge 0$ depending on the
direction of crossing of the origin.

A  simpler situation is obtained in the case of linear dispersion,
$e(p)=p$. There is now no ambiguity in the direction of crossing
and the energy-time phase space is again the plane. The map
$(q,p)\leftrightarrow (e,t)$ is, in fact, the identity
\begin{equation}
e=p,\quad t=-q.
\end{equation}
The usual coherent states are then also the coherent states on the
energy-time plane.

\subsection{Coherent states for linear dispersion}\label{csld}
The time-energy coherent states are
\begin{equation}\label{coherent}
\ket{t,e;\eps}= e^{i( tP+eX)} \ket{g_\eps},\quad [P,X]=-i,
\end{equation}
with $g_\eps$ Gaussian:
\begin{equation}\label{gaussian}
\brkt{p}{g_\eps}=\frac 1 {\root 4\of{\pi\eps^2}}\,
e^{-\frac{p^2}{2\eps^2}}.
\end{equation}
 They have the following properties \cite{perelomov}:
\begin{list}
{\Alph{mynr}}{\usecounter{mynr}
               \setlength{\rightmargin}{\leftmargin}} 
\item The states $\ket{t,e;\eps}$ are normalized.
\item $\ket{t,e;\eps}$ have Gaussian localization in 
time and 
energy near the point $(t,e)$ with width
$$\delta e\sim{\eps},\quad \delta t\sim
\frac 1 \eps, \quad \delta s\sim \frac \omega \eps.$$
Hence
$\omega$ plays the role of $\hbar$ in the epoch-energy plane.
\item\label{shift} $H_0$ is the generator of shifts of the coherent
states: 
$$
e^{-iH_0t'}\ket{t,e;\eps}=e^{-it'e/2}\,\ket{t-t',e;\eps}.
$$
\item The overlap of coherent states is:
$$\brkt{t,e,\eps }{t',e',\eps}=e^{-\frac{(e-e')^2}{4\eps^2}}
\,e^{-\frac{\eps^2(t-t')^2}{4}}e^{-i\frac{et'-e't}{2}}.$$ 
\item \label{resol} The coherent states give a resolution of the 
identity
$$\int \frac{dt \, de} {2\pi}\ket{t,e;\eps}\bra{t,e;\eps}=1.$$
\item The scalar product between coherent states and the eigenstates 
of 
$H_0=P$ is 
$$(E\ket{t,e;\eps }=e^{-ite/2}e^{-itE}\ 
\frac{e^{-(E-e)^2/2\eps^2}}{\root 4\of{\pi\eps^2}}.$$
\end{list}

\section{Scattering between channels with linear dispersion}
Linear dispersion approximates the low energy physics of electrons
in one dimensional channels provided the Fermi energy is large.
The price one pays is that the ``ultraviolet'' properties are
pathological. In particular, the spectrum is unbounded below and
this then leads to certain anomalies which must be correctly
interpreted. With linear dispersion one can also solve certain
models with interacting electrons \cite{ml}.

\begin{figure}[h]
\centering
\epsfxsize=3truein 
\epsfbox{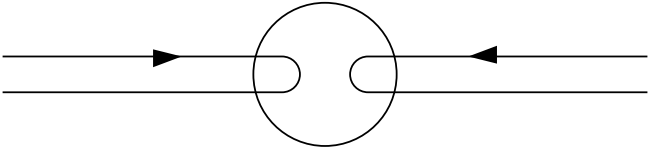}
\caption{A network of two channels. Each channel is chiral and lets 
particles propagate to and from infinity, according to the arrows. 
The circle denotes the
region where the channels are coupled. }\label{Network}
\end{figure}

In the following we shall study adiabatic scattering for {\it non
interacting} particles with linear dispersion. The particles move
on a collection of lines and are allowed to ``hop'' from one line
to the other and  scatter. Each line serves as an incoming and 
outgoing channel
since the flow on it is uni-directional. An example with two channels 
is shown in
Fig.~\ref{Network}. (Such models bear some resemblance to
Schr\"odinger operators on graphs \cite{smilansky}.) The Hilbert
space is  $\oplus_{j=1}^n L^2(\Real)$, a finite direct
sum. $j$ labels the scattering channels. $H_0$ is then
$$\big(H_0\psi\big)\,(x,j)=-i\,\psi'(x,j), \quad x\in \Real,\ 1\le 
j\le n.$$
For the interaction one may take, for example,
$$\bigg(\big(H(s)-H_0\big)\psi\bigg)(x,j)=\sum_{j'}v_{j,j'}(x,s)\psi(x,j')$$

with $v_{jj'}$ hermitian, and compactly supported. Alternatively, one 
may consider  finite rank perturbations.

\subsection{A soluble model\label{soluble}}
Here we describe a simple, time dependent, model for which the 
calculation of both the dynamical and frozen scattering matrices is 
reduced to quadrature. 

Consider scattering on the line with
$$H_0=P=-i\nabla,\quad H_s=P+f(s)\,V,\quad s=\omega t$$
with $\big(V\psi\big)(x)=v(x)\psi(x)$ a potential (multiplication 
operator) which is sufficiently regular and short range so that
$\int \abs{v(x)} \, dx,\ \int \abs{x v(x)} \, dx<\infty$ . The model 
has one channel and should not be confused with the 2-channel example 
pictured in Figure~\ref{Network}.

To calculate the dynamical S matrix note that 
\begin{equation}
\Omega(t,t'):=U(t,t')U_0(t'-t),
\end{equation}
satisfies the Volterra type equation:
\begin{equation}
\frac{\partial\Omega(t,t')}{\partial t'}=i f(\omega 
t')\Omega(t,t')V(t-t'), \quad 
\Omega(t',t')=1,
\end{equation}
with $V(t)$ the (backward) free Heisenberg evolution of the 
potential, i.e.
\begin{equation}
V(t):=U_0(t)\ V\ U_0(-t).
\end{equation}
Since $H_0=P$ is the generator of shifts, $V(t)$ is the shifted 
potential:
\begin{equation}
\big(V(t)\psi\big)(x)=v(x-t)\psi(x).
\end{equation}
In particular, $V(t)$ at different times commute, and the solution of 
the Volterra type problem is given simply by
\begin{equation}
\Omega(t,t')=e^{-i\int^{t-t'}_{0} f(s-\omega t'')V(t'')dt''},\quad 
s=\omega t.
\end{equation}
 From the definition of the wave operators based on time $s$, 
Eq.~(\ref{wave-operator}),  we obtain for the dynamical wave 
operators:
$$\Omega_-(s;H,H_0)=e^{-i\int_0^{\infty} f(s-\omega t') 
V(t')dt'},\quad 
\Omega_+(s;H,H_0)=e^{i\int_{-\infty}^0 f(s-\omega t') V(t')dt'}.$$
 From this we obtain for the dynamical scattering matrix 
\begin{equation}\label{ex-sd} S_d(s,H,H_0)=e^{-i\int^\infty_{-\infty} 
f(s-\omega t') V(t')\,dt'}.\end{equation}
The dynamical scattering matrix, as well as the wave operators, are 
local gauge transformations, i.e. multiplication by a {\em function} 
of position, of modulus one. 

The wave operators and the S matrix reduce to the frozen ones upon 
replacing 
the function $f(s-\omega t')$ by its frozen value $f(s)$, hence:
\begin{equation}\label{ex-sf}S_f(H_s,H_0)=e^{ 
-if(s)\,\int_{-\infty}^\infty V(t)\, dt}=e^{ 
-if(s)\,\int_{-\infty}^\infty V_+(t)\, dt},\end{equation} where 
$2 V_+(t)=V(t)+V(-t)$. $S_f$ is just a number, not a function of 
position. 

The frozen scattering matrix provide very little information on the 
potential $v(x)$, for it depends one just one number---the total 
weight of the potential\footnote{This is in sharp contrast with 
scattering 
problems where $H_0$ is the Laplacian \cite{inverse}.}. The 
dynamical S matrix, in contrast, provides independent information
about the potential for each value of $s$.

Since the frozen S matrix is independent of the incident energy, 
the Wigner time delay vanishes identically in this model: $\tau_w=0$. 
The 
(frozen) energy shift is just a real number (a multiple of the 
identity)
$$\E_f=\dot f(s)\int_{-\infty}^\infty v(x) dx.$$
In contrast, the dynamical energy 
shift, is the multiplication operator: 
\begin{equation}\E_d=\int^\infty_{-\infty}\dot f(s-\omega t') 
V(t')\,dt'.\end{equation} 

\subsection{The on-shell  scattering matrix and coherent states}
For later purposes we shall need the matrix elements of the frozen
S matrix. Since $\Sc_f$ commutes with $H_0$, the matrix elements
 are independent of $t$ and are related to the on-shell matrix by
\begin{eqnarray}\label{on-shell-coherent}
\bra{t,e,j;\eps }{\Sc_f\big(H_s,H_0\big)}\ket{t,e,j';\eps }&=&
\frac 1 {\sqrt \pi\eps }\,\int dE\, S_{jj'}(s,E) e^{-\frac{(E- 
e)^2}{\eps^2}}\nonumber \\ &=&
S_{jj'}(s, e)+O(\eps^2\partial_{EE}S).
\end{eqnarray}
The estimate is obtained by observing that 
$S_{jj'}(s,E)-S_{jj'}(s,e)$ does
not contribute to the integral to first order in $E-e$. Since
$$\left(\partial_{EE} S\right) S^\dagger= -\tau_w^2+i\tau^\prime_w,$$ 
(with prime denoting the derivative with respect to the energy) we 
see that the 
on-shell S matrix approximates the diagonal entries of the frozen S 
matrix, provided the Wigner time delay and its energy dependence are 
both small:
\begin{equation}\label{wigner}\eps^2(\tau_w^2+\abs{\tau'_w})\ll 
1.\end{equation} 

\section{The adiabatic time scale $\tau$}

In this section we compute, to leading order, the time scale $\tau$
relevant to adiabatic scattering. This time scale is defined so that 
$\omega\tau \ll 1$ characterizes the adiabatic regime in the 
sense that the frozen scattering data approximate the dynamical 
scattering
data. 

There are two results in this section, one positive and one
negative. The positive result says that, at least to leading order, 
$\tau$ can be  computed from
time independent quantities alone, Eq.~(\ref{tau+e}) below. The 
negative
result is that $\tau$ cannot be computed from the on-shell scattering
matrix and its derivatives. In particular, the Wigner time delay 
alone does not determine $\tau$.

Using 
Eqs.~(\ref{Sd0:Sds},\ref{S-h:h1},\ref{frozen-s}) and property 5.2.C
one finds
\begin{eqnarray} \label{remainder}
\bra{t,e,j;\eps }\big(\Sc_d(0;H,H_0)-\Sc_f(H_s,H_0)\big)\ket{t 
,e,j';\eps}=\\
\bra{0,e,j;\eps}\Omega^\dagger_+(H_s,H_0)
\big(\Sc_d(s;H,H_s)-1\big)\Omega_-(H_s,H_0)\ket{0,e,j;\eps}.\nonumber
\end{eqnarray}
The correction to the leading order of the S matrix 
can be approximated by an analog of the Born series \cite{yafaev}:
\begin{equation}\label{variation-s}
\Sc_d\big(s;H,H_s\big)-1\approx
-i\int_{-\infty}^{\infty}e^{iH_st'}\,\big(H_{s+\omega t'}-H_s\big)
\,e^{-iH_st'} \, dt'.\nonumber
\end{equation}
Since $H_{s+\omega t'}-H_s$ is supported near the origin, only
small $t'$ contribute to the matrix elements in 
Eq.~(\ref{remainder}). More precisely, this depends only on the time
localization property of either the bra or the ket. We can 
therefore
approximate $H_{s+\omega t'}-H_s\approx\omega t'\dot H_s$.
Using property 5.2.C
$$e^{-iH_st}\Omega_-(H_s,H_0)\ket{0,e,j;\eps}=
e^{-iet/2}\,\Omega_-(H_s,H_0)\ket{t,e,j;\eps}$$ we finally get
\begin{eqnarray} \label{remainder-leading}
\bra{t,e,j;\eps } \big(\Sc_d(0;H,H_0)-\Sc_f(H_s,H_0)\big)\ket{t
,e,j';\eps}\approx -i\omega\tau(e,s;\eps )
\end{eqnarray}
where
\begin{equation}\label{tau+e}
\tau(e,s;\eps )=\int_{-\infty}^\infty \bra{t',e,j;\eps }
\Omega^\dagger_+(H_s,H_0)\dot H_s\Omega_-(H_s,H_0)
\ket{t',e,j;\eps}\,t'\,dt'.
\end{equation}
 $\tau(e,s;\eps )$
involves the frozen wave operators and the rate of change of the 
Hamiltonian at the epoch $s$. In particular, one can
use methods of time-independent scattering theory to compute it. 
It is in general complex. The adiabatic time scale, $\tau=\abs{\tau 
(e,s;\eps
)}$ is a measure of the error. $\omega\tau\ll 1$ then clearly 
characterizes the adiabatic regime.

Propagation estimates can, and have been, used \cite{schnee} 
to bound the  error in the frozen data. These estimates yield 
bounds on $\tau$.

\subsection{Example: the soluble model.} 
For the case of one channel scattering with 
$H(s)=P+f(s)V$, by Eq.~(\ref{ex-sd},\ref{ex-sf})\begin{eqnarray}
S_d(s;H,H_0)-S_f(H_s,H_0)&=&\left(e^{- i\int^\infty_{-\infty} 
\big(f(s-\omega t)-f(s)\big) V(t)\,dt}-1\right)\ 
S_f(H_s,H_0)\nonumber \\
&\approx& i\omega \dot f(s)\left(\int^\infty_{-\infty} t 
V(t)\,dt\right)\, S_f(H_s,H_0).
\end{eqnarray}
The adiabatic time scale $\tau$ is, in analogy with 
Eq.~(\ref{remainder-leading}), the multiplication operator:
\begin{eqnarray}
&&\tau\approx -\dot f(s)\left(\int^\infty_{-\infty}\!\! t 
V(t)\,dt\right)\, 
S_f(H_s,H_0)=
-\dot f(s)\left(\int^\infty_{-\infty}\!\! t V_-(t)\,dt\right)\, 
S_f(H_s,H_0);\\
&& \ \ \ \ \ \ \ \ \ \ \ \ \ \ \ \ \ \ \ \ \ \ \ \
2 V_-(t)=V(t)-V(-t).\nonumber\end{eqnarray}
By Eq.~(\ref{ex-sf}) the frozen S matrix  only depends on $V_+$, 
while the error only depends on $V_-$. Since $V_-$ and $V_+$ are 
independent this shows that the error term
in the adiabatic expansion cannot be estimated in terms on the
frozen scattering data alone. 

Combining Eqs.~(\ref{on-shell-coherent}) and (\ref{remainder-leading})
we obtain a relation between matrix elements of the dynamical S 
matrix and the on-shell S matrix: 
\begin{equation}\label{combined}
\bra{t,e,j;\eps }\Sc_d(0;H,H_0)\ket{t,e,j';\eps}=S_{jj'}(s, e)
+O\big(\eps^2(\tau_w^2+\abs{\tau'_w})+\omega\tau(e,s;\eps)\big),
\end{equation} 

\section{The energy shift}

The energy shift is a first order quantity, nevertheless, it is
determined, to leading order, by the frozen data:

\begin{eqnarray}
\quad\quad\braket{t,e,j;\eps}{\E_d(0)}{t,e,j';\eps} &\approx&
i\,\big(\dot S(s,e) S^\dagger (s,e)\big)_{jj'},\quad
(s=\omega t). \label{e-thawed-frozen}
\end{eqnarray}
We first remark that (\ref{remainder-leading},\ref{tau+e}) generalize 
to
off-diagonal matrix elements, i.e., the time $t$ in the ket 
$\ket{t,e,j';\eps}$ may be shifted to $t+\Delta t$ (resp. $t'+\Delta 
t$ in
(\ref{tau+e})) while leaving the bra
unchanged. By the translation property of coherent states, 
property 5.2.C,
multiplication by $H_0$ can be traded for derivative with respect
to time. Hence
\begin{eqnarray}
\lefteqn{
\bra{t,e,j;\eps}\big[ H_0,\Sc_d(0;H,H_0)\big]\ket{t+\Delta 
t,e,j';\eps}}
\nonumber\\
&=&\!\! i\partial_t\,
\bra{t,e,j;\eps}\Sc_d(0;H,H_0)\ket{t+\Delta t,e,j';\eps}\nonumber
\\
\!\! &\approx& \!\! i\partial_t\,
\bra{t,e,j;\eps}\Sc_f(H_s,H_0)\ket{t+\Delta t,e,j';\eps}\nonumber\\
\!\! &=&i\omega\bra{t,e,j;\eps}\dot\Sc_f(H_s,H_0)\ket{t+\Delta 
t,e,j';\eps}.
\label{eq:comm}
\end{eqnarray}
In principle, the order of the error in the frozen data in the passage 
from the second to the third line does not determine 
the order of the error in derivatives, but this can be justified in 
the present case. The last identity in the equation above can be seen from 
\begin{eqnarray}
\lefteqn{
\bra{t,e,j;\eps }{\Sc_f\big(H_s,H_0\big)}\ket{t+\Delta t,e,j';\eps }}
\nonumber\\
&=&
\frac 1 {\sqrt \pi\eps }\,\int dE\, S_{jj'}(s,E) e^{-\frac{(E- 
e)^2}{\eps^2}}e^{-i(\Delta t)e/2}e^{-i(\Delta t)E}.\nonumber
\end{eqnarray}
We then multiply (\ref{eq:comm}) with the complex conjugate of the 
mentioned generalization of 
(\ref{remainder-leading}) and integrate over $\Delta t$ using 
property 
5.2.E. The result then heuristically follows from 
Eq.~(\ref{dotS}) and the statement for $\E_f$ analogous to 
(\ref{on-shell-coherent}).\footnote{ An alternate derivation of 
\ref{e-thawed-frozen} can be made, more 
directly, starting  from  the rhs of Eq. \ref {dotS} and using Born's  
expansion.}

The energy shift plays a role in the theory of adiabatic quantum
pumps. In particular, the pumped charge, the entropy production and
noise generation in quantum pumps can all be expressed in terms of the
energy shift \cite{aegs}. It is remarkable that basic properties of
adiabatic quantum pumps can be understood, to leading order, in terms
of the frozen scattering data alone.




\bibliographystyle{amsplain}

\end{document}